\documentclass[aps,pra,preprint]{revtex4}
\usepackage{graphics}
\usepackage{latexsym}
\usepackage{amssymb}
\usepackage{epsfig}
\usepackage[dvips]{color}
\usepackage{epstopdf}

\begin{document}
\hspace {-8mm} {\bf\LARGE Is it possible to construct excited-state energy functionals by splitting k-space?} \\ \\

\hspace {-8mm} {\bf M.Hemanadhan and Manoj K. Harbola}\\
{Department of Physics, Indian Institute of Technology,
Kanpur 208016, India}

\begin{abstract}
We show that our procedure of constructing excited-state energy functionals by splitting k-space, employed 
so far to obtain exchange energies of excited-states, is quite general.  We do so by applying the same 
method to construct modified Thomas-Fermi kinetic energy functional and its gradient expansion up to the 
second order for the excited-states.  We show that the resulting kinetic energy functional has the same 
accuracy for the excited-states as the ground-state functionals do for the ground-states.
\end{abstract}

\maketitle
\section {Introduction}
Since the inception of ground-state density functional theory (DFT)\cite{hk,ks,parr,grossd,march},
efforts have been made to extend it to excited-states.  Such attempts include the work of Ziegler et al.
\cite{ziegler}, Gunnarsson et al. \cite{gunnar}, von Barth \cite{von}, Perdew and Levy  \cite{levper},
Pathak \cite{pathak}, Theophilou \cite{theophi}, Oliveira, Gross and Kohn \cite{grossolkohn,olgrosskohn},
Nagy \cite{nagy}, Sen \cite{sen} and Singh and Deb \cite{debsingh}.  However, a general excited-state
density functional theory for individual excited-states, akin to its ground-state counterpart, has started
taking shape \cite{harb1,harb2,levynag,gor,samal2,samalh,hsamal,aip,ayers} only over the past decade or so.

In density functional theory, energy of a system is expressed as a functional of the density of the system.
History \cite{mkhab} of writing energy of a system in terms of its density is as old as quantum-mechanics 
itself.  In an attempt to simplify the problem of interacting electrons, Thomas \cite{thomas} and 
Fermi \cite{fermi} expressed the kinetic energy of a many electron system approximately by employing the 
expression \cite{kittel} for the kinetic energy of the homogeneous electron gas (HEG).  Similarly, 
Dirac \cite{dirac} gave an
approximate expression for the exchange energy of a many-electron system by employing the corresponding
HEG formula. With the Hohenberg-Kohn \cite{hk} discovery of the one-to-one map between the ground-state
density and the Hamiltonian of a system, it became clear that the energy of a system can indeed be expressed
as a functional of its ground-state density; however, the functional is not known exactly.  In the
Kohn-Sham formulation \cite{ks} of DFT, the kinetic energy component of the total energy is treated highly
accurately by writing it in terms of orbitals of an auxiliary system. Thus the non-interacting kinetic energy is expressed in terms of the Kohn-Sham orbitals $\left | \phi_{i} \right \rangle $ as (atomic units are used throughout the paper so that we take $\hbar=m_e=|e|=1$)  
\begin{equation}
 \sum_{i} f_{i} \left \langle \phi_{i}   \right|  -\frac{1}{2}  \nabla^2  \left | \phi_{i} \right \rangle
\label{ke}
\end{equation}
  where  $f_{i} $ represent the occupation of $i^{th}$ orbital. For the ground-state, $f_i=1$ for $i^{th}$ orbital of each spin if $i\leq i_{max}$ where $i_{max}$ is the index of the uppermost occupied orbital and is $0$ for all the higher orbitals.  For an excited-state, the occupation is different from the ground-state; for example it could be equal to $1$ for $i\leq i_1$, $0$ for $i_1 < i\leq i_2$ and $1$ again for $i_2 < i\leq i_3$, as shown schematically in Fig. 1.  On the other hand, the exchange and correlation energies  are still expressed 
approximately in terms of the density. Foremost among these approximations are the local-density
approximation (LDA) and the local spin-density approximation (LSD).  In these approximations, the exchange
and correlation energies are expressed in terms of the density by employing the corresponding expression
for the HEG.  Thus the LDA for the exchange energy is the same as the Dirac expression for it.  Over the
years, far more accurate functionals \cite{gga} for exchange and correlation energies have been constructed
by going beyond the LDA and including corrections in terms of the gradient of the density.  The leading
term in most of these functionals is the LDA/LSD functional and in the limit of the gradient of
the density vanishing, the functionals indeed reduce to the latter.

Given this background, a question that arises naturally in the development of excited-state DFT is if
it would possible to construct energy functionals for these states with similar accuracy as is obtained
in the ground-state functionals. In particular it is important to develop an LDA functional for the
excited-states since that is the foundation on which more accurate functionals are built. We have
recently constructed an exchange energy functional for excited-states within the LDA.  This has been
done by splitting the k-space in accordance to the occupied and unoccupied orbitals of the excited-state,
as shown in Fig. 1.  In the figure, we have some orbitals - the core orbitals - including the lowest
energy orbitals that are occupied, then some empty orbitals and then some more orbitals - the shell
orbitals - that are occupied again.  The k-space, also shown in the figure, is accordingly split such
that it is occupied from $k=0$ to $k=k_1$, empty from $k_1$ to $k_2$ and then occupied again from $k_2$
to $k_3$.  Here $k_1$, $k_2$ and $k_3$ are given by the equations 

\begin{figure}
\begin{center}
\includegraphics[width=5in,height=3.0in]{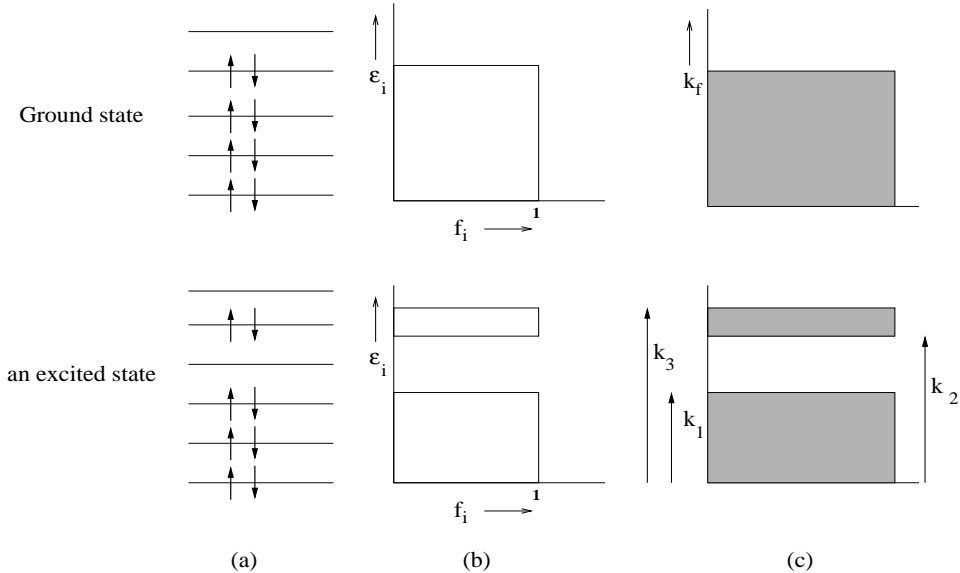}
\caption{ Orbital occupation of electrons (a) and the corresponding $f_i$ for each spin drawn continuously as a function of orbital energy $\epsilon_i$ (b) for the ground and an excited-state of a finite system. The corresponding k-space occupation (c), in the ground and an excited state configuration similar to that shown in (a) for a homogeneous electron gas (HEG).
}
\end{center}
\end{figure}

\begin{equation}
k_{1}^{3}(\textbf{r}) = 3\pi^{2}\rho_{c}(\textbf{r})
\label{eq:k1}
\end{equation}
\begin{equation}
k_{2}^{3}(\textbf{r})-k_{1}^{3}(\textbf{r}) = 3\pi^{2}\rho_{vac}(\textbf{r})
\label{eq:k2}
\end{equation}
\begin{equation}
k_{3}^{3}(\textbf{r})-k_{2}^{3}(\textbf{r}) = 3\pi^{2}\rho_{s}(\textbf{r})
\label{eq:k3}
\end{equation}
where $\rho_{c}$ and $\rho_{s}$ are the electron densities corresponding to the core and the shell
orbitals. Similarly, $\rho_{vac}$ is the electron density corresponding to the set of unoccupied
orbitals. 
Thus \\
\begin{center}
$\rho_{c}(\bf{r})$ = $\sum_{i}{\left| \phi_{i}^{core}(\bf{r})\right|}^{2}$ \\                                                \end{center}
\begin{center}
$\rho_{vac}(\bf{r})$ = $\sum_{i}{\left| \phi_{i}^{unocc}(\bf{r})\right|}^{2}$ \\                                           \end{center}
\begin{center}
$\rho_{s}(\bf{r})$ = $\sum_{i}{\left| \phi_{i}^{shell}(\bf{r})\right|}^{2}$ \\                                                \end{center}

The total electron density $\rho({\bf r})$ is given as
\begin {equation}
\rho({\bf r})=\rho_{c}({\bf r})+\rho_{s}({\bf r})
\label{eq:den}
\end {equation}
For detailed derivation of these equations, we refer the reader to the next section.  Employing the
exchange energy functional developed by us, we have been performing accurate calculations
\cite{samal2,samalh,hsamal,aip} of excited-state energies of a variety of systems including the
band gaps \cite{rahman} of a wide variety of semiconductors in the recent past.

Although the results obtained by us with the excited-state exchange energy functional are impressive, the
question that we have been asking ourselves is if the method employed by us - that of splitting the
k-space - to construct the functional is general.  If the answer is in the affirmative, the same method
should also lead to reasonably accurate functionals for other components, viz. the kinetic and the correlation energies,
of the total energy. Further, we should be able to build on the LDA to include higher order corrections in
terms of the gradient of the density.  In this paper we address this question in connection with the
non-interacting kinetic energy of a system of electrons. Our aim in these investigations is to explore if a kinetic energy functional for excited states, constructed by splitting the k-space, gives similar accuracy for exact kinetic energy of these states as the well known Thomas-Fermi or the gradient-expansion approximation (GEA) functionals\cite{parr,grossd} do for the ground states.  We show in this paper that it does.  Thus our present results demonstrate the robustness of our
procedure of constructing energy functionals for the excited-states of a many-electron system.

We note that besides the GEA functional, there is another approach to constructing kinetic-energy
functionals \cite{gazquez} for the ground-state, which employs two exact asymptotic forms: Thomas-Fermi
for the HEG and the von-Weizsacker term \cite{vonW} for one-orbital systems.  If our approach has
universality, it should also work for functional such as the G\'{a}zquez-Robles functional proposed in reference \cite{gazquez}.  We show in this paper that it does.

We start in the next section with a description of the Thomas-Fermi approximation for the non-interacting
kinetic energy for the ground-state.  This approximation is the LDA for the kinetic energy.  We then discuss
the gradient expansion approximation (GEA) for the kinetic energy up to the second-order in the
density gradient.  Results for a few atomic systems and the key features of these results are then
discussed.  This forms the background against which the kinetic energy functional for the excited-state
is then constructed and tested in the section after the next one.  We end the paper with some concluding
remarks.

We point out that our aim in this paper is to explore conceptually if our approach yields kinetic energy functionals that have accuracy similar to their ground-state counterparts.  Our work shows that it does.  Question arises: Can these functionals be applied to obtain average excited-state energies.  This possibility is being explored.  However, density-based functionals cannot be expected to reproduce the exact answer for the kinetic energy, as given by equation 1.  Therefore the operational  utility of the  kinetic energy density-functionals for excited-states is similar to that of traditional Thomas-Fermi functional or its extensions for the ground-state.

\section {LDA and GEA up to the second order for the non-interacting kinetic energy of the ground-state}
The basis of the LDA is the homogeneous electron gas for which the kinetic and the exchange energies can
be expressed in a rather simple form involving the density of the system.  For the non-interacting kinetic
energy we consider a gas of non-interacting electrons that fill the k-space from $k=0$ to $k=k_F$ because
of the Pauli exclusion principle.  The wavefunction for an electron in a state specified by wavevector
${\bf k}$ is
\begin{equation}
\psi_{\bf k}({\bf r}) = \frac{1}{\surd V}exp(\imath{\bf k.r}),
\label{5}
\end{equation}
where $V$ is the volume over which the periodic boundary conditions are applied on the wavefunction.
Assuming the volume to be a large cube of side $L$, the wavevectors ${\bf k}$ take the values
\begin{equation}
{\bf k} = \frac{2\pi}{L}(n_1\hat{\bf x}+n_2\hat{\bf y}+n_3\hat{\bf z})
\label{6}
\end{equation}
where $n_i=1, 2, 3\ldots$ with the maximum value such that the magnitude of the largest ${\bf k}$ is
$k_F$. The density of k-points in the k-space is therefore $\frac{V}{8\pi^3}$ and the density of states
including the spin of the electrons is $\frac{V}{4\pi^3}$.  Equating the total number of electrons $N$
within the volume $V$ to the number of states within a sphere of radius $k_F$, referred to as the Fermi
sphere, leads to
\begin{equation}
k_F=(3\pi^2\rho)^{\frac{1}{3}}
\label{eq:kf}
\end{equation}
where $\rho=\frac{N}{V}$ is the number density of the homogeneous electron gas. Similarly the total
kinetic energy is calculated by summing the kinetic energy $\frac{k^2}{2}$ of a state specified
by the wavevector ${\bf k}$ over the Fermi sphere. It gives the kinetic energy density or the
kinetic energy $t_s$ per unit volume to be
\begin{eqnarray}
t_s &=& \frac{k^5_F}{10\pi^2} \nonumber \\
          &=& \frac{3}{10}(3\pi^2)^{\frac{2}{3}}\rho^{\frac{5}{3}}
\label{8}
\end{eqnarray}

The local density approximation to the kinetic energy $\sum_{i} f_{i} \left \langle \phi_{i}   \right|  -\frac{1}{2}  \nabla^2  \left | \phi_{i} \right \rangle $   for the ground-state of an inhomogeneous electron
gas, such as that in an atom or a molecule, of space-dependent density $\rho({\bf r}) = \sum_{i} f_{i} \left | \phi_i({\bf r}) \right |^2 $ corresponds to
approximating the kinetic energy density at each point by the formula above and integrating it over the
entire volume.  This leads to the Thomas-Fermi kinetic energy functional
\begin{eqnarray}
T^{(0)}_s[\rho] &=& \frac{1}{10\pi^2}\int k^5_F({\bf r}) d{\bf r} \nonumber \\
                &=& \frac{3}{10}(3\pi^2)^{\frac{2}{3}}\int\rho^{\frac{5}{3}}({\bf r})d{\bf r}
\label{eq:tf}
\end{eqnarray}
where $k_F({\bf r})$ and $\rho({\bf r})$ at each point in space are related by equation~\ref{eq:kf}.  In equation~\ref{eq:tf}
the superscript $(0)$ indicates that this is the zeroth-order approximation to the exact kinetic energy for
an inhomogeneous electron gas.  It is well known to underestimate the exact kinetic energy.
If the number of up and down spin electrons is different, the functional given above can be written in
terms of the spin densities $\rho\uparrow$ and $\rho\downarrow$ as
\begin{equation}
T[\rho\uparrow,\rho\downarrow] = \frac{1}{2}(T^{(0)}_s[2\rho\uparrow]+T^{(0)}_s[2\rho\downarrow])
\label{eq:lsd}
\end{equation}
Exact kinetic energy for some closed-shell hydrogen-like atoms and the Thomas Fermi approximation to it
for the same atoms is given in Table I. Table II gives the exact kinetic energy for atoms from H to Ne for
the density obtained by solving the Kohn-Sham equation for it within the Gunnarsson-Lundquist
parametrization \cite{gunnar} of the LDA for the exchange-correlation energy.  As is evident from the
Tables, Thomas-Fermi functional underestimates the exact kinetic energy by about $5\%$ to $10\%$.

The first correction to the Thomas-Fermi functional in terms of the density gradient is
proportional to the square of the gradient of the density and is given as \cite{parr,grossd}
\begin{equation}
T^{(2)}_s[\rho] = \frac{1}{72}\int\frac{|\nabla\rho({\bf r})|^2}{\rho({\bf r})}d{\bf r}
\label{eq:gea2}
\end{equation}
This is easily derived \cite{hk,rmp} from the expansion of the response function of a non-interacting
electron gas.  The correction term is also generalized in terms of the spin densities as given by
equation \ref{eq:lsd}.  Equation \ref{eq:gea2} represents the lowest-order gradient correction to the
Thomas-Fermi functional.  The gradient-corrected kinetic energy $T^{(0)}+T^{(2)}$ is also given in
Tables I and II for the atomic systems given there. It is seen that the inclusion of the second-order
correction brings the approximate kinetic energy closer to the exact one, with the difference being
less than $1\%$.  The question that we now address is if kinetic energy functionals can also be written
for excited states using ideas employed to generate functionals for the ground-state.

\section {LDA and GEA up to the second order for the non-interacting kinetic energy of excited-states}
One straightforward choice for the functionals to be employed for excited-states is to use the
ground-state functionals described in the section above.  However, the way k-space is occupied
to construct the ground-state functionals does not reflect proper occupation of orbitals that are
occupied in an excited-state. While for the excited-states of a given number of electrons in a
homogeneous electron gas we expect orbitals with relatively larger magnitude of wavevectors to be
occupied, this does not happen if we use the ground-state functional to approximate the kinetic energy
of excited-states; as such the ground-state functional would underestimate the kinetic energy of
excited-states by a much larger amount than the proper excited-state functional should. This is shown
in Table III where we have shown the approximate kinetic-energy, calculated using the ground-state LSD
and GEA functionals of the section above, of some excited-states of a few hydrogen-like atoms and
have compared them to the exact kinetic energy.  It is seen from the Tables that the ground-state
functionals indeed underestimate the exact kinetic energy of excited states by significantly larger
amount than they do for the ground-states.  In Table IV, the approximate kinetic energies are compared
to the exact kinetic energies for excited-state densities obtained by solving the Kohn-Sham equation with
the Gunnarsson-Lundquist parametrization. Here also the error in numbers obtained by applying the
ground-state functionals is quite large.

To develop a kinetic energy functional corresponding to excited-states, we have proposed \cite{samalh}
in the context of exchange energy that the k-space be split in accordance to the occupation of orbitals
in the excited-state of a system. This is shown in Fig. 1 for an excited-state where some lowest lying
orbitals - the core orbitals - are occupied, then there are some vacant orbitals and then some more
orbitals - the shell orbitals - are occupied.  According to our method of constructing excited-state
functionals for such a system, the k-space is also occupied correspondingly with orbitals up to $k_1$
being occupied with $k_1$ given by equation~\ref{eq:k1}, orbitals between $k_1$ and $k_2$ being vacant
with $k_2$ given by equation ~\ref{eq:k2}, and again orbitals from $k_2$ to $k_3$ being occupied with
$k_3$ given by equation ~\ref{eq:k3}.  Now steps leading to equation~\ref{eq:tf} are taken to derive
the kinetic energy for such a system.  The corresponding LDA functional $T^{*(0)}$ for the excited-state 
is then given as
\begin{equation}
T^{*(0)}(k_1,k_2,k_3)=\frac{1}{10\pi^2}\int \left(k^5_1({\bf r})+k^5_3({\bf r})-k^5_2({\bf r})
\right)d{\bf r}
\label{eq:exke0}
\end{equation}
The spin-density generalization of equation~\ref{eq:exke0} is given by equation~\ref{eq:lsd}.
We call the functional given by equation~\ref{eq:exke0} the modified Thomas-Fermi functional. In the functional above the term $\frac{1}{10\pi^2}\int \left(k^5_1({\bf r})\right)d{\bf r}$ represents the kinetic energy of the core orbitals whereas $\frac{1}{10\pi^2}\int \left(k^5_3({\bf r})-k^5_2({\bf r})\right)d{\bf r}$ that of the shell orbitals. Thus $\sum_{i}  \left \langle \phi_{i}^{core}   \right|  -\frac{1}{2}  \nabla^2  \left | \phi_{i}^{core} \right \rangle $  is approximated by the former while $\sum_{i}  \left \langle \phi_{i}^{shell}   \right|  -\frac{1}{2}  \nabla^2  \left | \phi_{i}^{shell} \right \rangle $ by the latter.

We now test the functional of equation~\ref{eq:exke0} for excited-states of hydrogen-like and real atoms.  
Shown in Table III
are the approximate kinetic energies calculated using the functional of equation~\ref{eq:exke0} for
excited-states of hydrogen-like atoms.  The numbers shown are for excited-states in which orbitals
up to principal quantum number $n_1$ are occupied, those from $n_1+1$ to $n_2$ are vacant and than $n_2+1$ to $n_3$
are again occupied.  It is seen that the energies calculated with the functional
of equation~\ref{eq:exke0} are better approximation to the exact kinetic energy in comparison to
the ground-state functional of equation~\ref{eq:tf}. Thus while the ground-state kinetic energy functional
$T^{(0)}$ given by equation \ref{eq:tf} underestimates the excited-state kinetic energy by a
substantial amount, the excited-state functional $T^{*(0)}$ given by equation \ref{eq:exke0} has
the same accuracy for the excited-states as the ground-state functional does for the ground-state.
In Table IV the numbers for approximate kinetic energy as obtained by applying the Thomas-Fermi
functional and the modified Thomas-Fermi functional of equation~\ref{eq:exke0} for the density of
Kohn-Sham are given.  These densities have been obtained by solving the Kohn-Sham equation within the
local-spin density approximation for the exchange-correlation energy .  The numbers are compared with
the exact kinetic energy $\sum_{i} f_{i} \left \langle \phi_{i}   \right|  -\frac{1}{2}  \nabla^2  \left | \phi_{i} \right \rangle$ with $f_i$ representing the number of electrons in the $i^{th}$ orbital , obtained from the Kohn-Sham orbitals $\left | \phi_{i} \right \rangle $.  Similar to the case of hydrogen-like
atoms, here too the functional of equation~\ref{eq:exke0} gives kinetic energies that are better than those obtained from the ground-state functional and has similar accuracy as the ground-state
functional does for the ground-state densities.  We thus conclude that the modified Thomas-Fermi
functional of equation \ref{eq:exke0} is the correct zeroth-order approximation for the kinetic
energies of excited-states. More importantly, this indicates that the idea of constructing
excited-state energy functionals by splitting the k-space is a sound one.

We next discuss the gradient expansion approximation for the excited-states kinetic energy. Since the
kinetic energy is a sum of kinetic energy of individual orbitals, the second order correction to the
kinetic energy for an excited-state can also be written exactly in the same manner as the zeroth order
approximation given by equation \ref{eq:exke0}. Thus the second-order gradient correction to the
excited-state kinetic energy is given as
\begin{equation}
T^{*(2)}(k_1,k_2,k_3)= \frac{1}{72}\int\frac{|\nabla\rho({\bf r};k_1)|^2}{\rho({\bf r};k_1)}d{\bf r}
                     + \frac{1}{72}\int\frac{|\nabla\rho({\bf r};k_3)|^2}{\rho({\bf r};k_3)}d{\bf r}
                     - \frac{1}{72}\int\frac{|\nabla\rho({\bf r};k_2)|^2}{\rho({\bf r};k_2)}d{\bf r}
\label{eq:exke2}
\end{equation}
Like in the functional of equation~\ref{eq:exke0}  the term $\frac{1}{72}\int\frac{|\nabla\rho({\bf r};k_1)|^2}{\rho({\bf r};k_1)}d{\bf r}$ gives the gradient correction to the core-orbitals kinetic energy while the last two terms give it for the shell orbitals.
Here $\rho({\bf r};k)=\frac{k^3}{3\pi^2}$ is the ground-state density corresponding to Fermi
wavevector $k$. The spin-density generalization of equation
\ref{eq:exke2} is given by equation \ref{eq:lsd}. In Tables III and IV we also show the second-order
corrected kinetic energy $T^{*(0)}+T^{*(2)}$ for excited-states of hydrogen-like and real atoms, respectively, and compare it to the exact
kinetic energies. It is again seen that the second order-correction calculated by using
equation~\ref{eq:exke2} leads to improved kinetic energies for the excited-states.

As an extreme test for the functional of equations \ref{eq:exke0} and \ref{eq:exke2}, we apply
them to excited-states where there are no core electrons, i.e. all the electrons have been excited.
The exact and approximate kinetic energies for such states are given in Table V for the Kohn-Sham
densities. Comparison of the numbers given shows the following:  while the ground-state functional
of equations \ref{eq:tf} and \ref{eq:gea2} underestimate the exact kinetic energy by very large
amount, the excited-state functionals of equations \ref{eq:exke0} and \ref{eq:exke2} bring the
error down significantly.  This again points to the soundness of the idea - that of splitting
the k-space - behind the construction of these functionals.

\section{G\'{a}zquez-Robles functional for excited-states}
As mentioned in the introduction, there are other forms of the kinetic energy functional for the ground-state that are based on considerations other than the LDA and it gradient expansion.  One of these approaches constructs a functional by combining the von-Weizsacker functional
\begin{equation}
 T^{W}[\rho]= \frac{1}{8}\int\frac{|\nabla\rho({\bf r}|^2}{\rho({\bf r}}d{\bf r},
\label{eq:tvonW}
\end{equation}
which is exact for one-orbital systems and the Thomas-Fermi (equation \ref{eq:tf}) functional with a correction factor
\begin{equation}
 C(N)=\left(1-\frac{2}{N}\right)\left(1-\frac{A_1}{N^{\frac{1}{3}}}+\frac{A_2}{N^{\frac{2}{3}}}\right),
\end{equation}
where $N$ is the number of electrons in the system.  Thus the final functional is
\begin{equation}
 T^{(0)g}[\rho]=T^{W}(\rho)+C(N)T_s^{(0)}(\rho)
\label{eq:tgazq}
\end{equation}
The constants $A_1=1.314$ and $A_2=0.0021$ for spin-compensated case \cite{grossd}.  It  is easily generalized to the 
spin dependent case through equation \ref{eq:lsd}.  In the functional above, the von-Weizsacker term gives accurate kinetic energy for the lowest orbital and the contribution from the rest of the orbitals is accounted for by the second term.  Thus the factor $\left(1-\frac{2}{N}\right)$ in the second term plays an important role of subtracting from the Thomas-Fermi functional the kinetic energy contribution of the
lowest orbital, treated exactly by the first term.

Applying the same arguments that were used to derive
equation \ref{eq:exke0} and \ref{eq:gea2} - that the kinetic energy for an excited-state is written as
a combination of the ground-state kinetic energy functionals corresponding to the wavevectors $k_1$, $k_2$ and $k_3$ - we write the excited-state G\'{a}zquez-Robles functional as 
\begin{equation}
 T^{*g}(k_1,k_2,k_3)=T^{(0)g}(\rho({\bf r};k_1))+T^{(0)g}(\rho({\bf r};k_3))-T^{(0)g}(\rho({\bf r};k_2))
\label{eq:exgazq}
\end{equation}
We have also tested the ground-state G\'{a}zquez-Robles functional (equation 17) and its excited-state generalization (equation 18) for the excited-states studied in Tables IV and V.  The results are shown in Tables VI and VII.  It is evident from the numbers presented that with the G\'{a}zquez-Robles functional also, our approach leads to an excited-state functional that estimates the kinetic energy of an excited-state better than its ground-state counterpart.  We note, however, that unlike the GEA functional the G\'{a}zquez-Robles functional is not uniformly accurate for all the excited-states studied.  This could be because the parameters of the functional have been optimized using the
ground-state kinetic energies of atoms within the Hartree-Fock theory.  Nonetheless, by applying our approach to two kinetic-energy functionals, which are derived by two different methods, we have shown that our method leads to improved functionals for excited states.

\section {Concluding remarks}
In this work we have tested the idea of constructing the LDA to excited-state energy functionals of time-independent density functional theory by splitting the k-space in the context of non-interacting kinetic energy functionals.
Our results show that the functionals obtained by such a method have the same accuracy for the 
excited-states as the ground-state functionals do for the ground-states. Further, we have shown that 
gradient correction can also be made on such functionals.  The general nature of our proposal is evident from the fact that applying it to a different kinetic-energy functional also leads to an improved functional for the excited-states.  In the future we would like to derive the
gradient correction given by equation~\ref{eq:exke2} in a manner similar to that~\cite{rmp} for
the ground-state, i.e. from the response function of the excited HEG. Further, it would also
be interesting to see if excited-state functional derived here can be used to approximately
calculate excited-state energies by employing a variational form for the excited-state densities.

{\bf Acknowledgment:} We thank Md. Shamim for useful discussions.  We also than one of the anonymous referees for suggesting that we test our method on functionals other than the GEA functional.

\newpage

\begin{table}
\caption{Closed Shell : Exact and Thomas Fermi (Equation~\ref{eq:tf} of the text ) $T^{(0)}$ and Gradient corrected kinetic energy
$T^{(0)}+T^{(2)}$ (Equation \ref{eq:gea2} of the text ) for
closed shell hydrogen like atoms. Numbers
given are in atomic units. Percentage errors as shown in brackets under each number}
\vspace{0.2in}
\begin{tabular}{p{3cm}p{2cm}p{3cm}p{3cm}}
 \hline
atoms&$T^{Exact}$&\hspace{0.3cm}$T^{(0)}$&
$T^{(0)}+T^{(2)}$\\
\hline
$He(1s^{2})$ &4&3.672&4.134\\
&&(8.2)&(3.4)\\
$Be([He]2s^{2})$ &20&17.719&19.785\\
&&(11.4)&(1.1)\\
$Ne([Be]2p^{6}\;)$ &200&188.849&202.869\\
&&(5.6)&(1.4)\\
$Mg([Ne]3s^{2})$ &304&284.712&305.978\\
&&(6.3)&(0.6)\\
$Ar([Mg]3p^{6}\;)$ &792&737.963&790.652\\
&&(6.8)&(0.2)\\
\hline
\end{tabular}
\end{table}

\begin{table}
\caption{Exact Kinetic Energy of ground state of some atoms as obtained by solving the Kohn-Sham equation 
with the Gunnarsson-Lundquist parametrization of the LSD for exchange and correlation energy. Numbers
given are in atomic units. The exact kinetic energy is compared with the Thomas-Fermi (Equation~\ref{eq:tf}) 
and gradient corrected functional (Equation~\ref{eq:exke0})}
\vspace{0.2in}
\begin{tabular}{p{3cm}p{2cm}p{3cm}p{3cm}}
\hline
atoms& KS   & $T^{(0)}$ &
 $T^{(0)}+T^{(2)}$\\
\hline
$H(1s^{1})$ &0.430&0.390&0.438\\
&&(9.3)&(1.8)\\
$He(1s^{2})$ &2.780&2.468&2.777\\
&&(11.2)&(1.1)\\
$Li([He]2s^{1})$ &7.269&6.521&7.305\\
&&(10.3)&(0.5)\\
$Be([He]2s^{2})$ &14.331&12.860&14.347\\
&&(10.3)&(0.1)\\
$B([Be]2p^{1})$ &24.201&21.649&24.040\\
&&(10.5)&(0.7)\\
$C([Be]2p^{2})$ &37.277&33.476&36.980\\
&&(10.2)&(0.8)\\
$N([Be]2p^{3})$ &53.899&48.946&53.778\\
&&(9.2)&(0.2)\\
$O([Be]2p^{4})$ &74.223&67.084&73.406\\
&&(9.6)&(1.1)\\
$F([Be]2p^{5})$ &98.742&89.450&97.472\\
&&(9.4)&(1.3)\\
$Ne([Be]2p^{6})$ &127.794&116.838&126.778\\
&&(8.6)&(0.8)\\
\hline

\end{tabular}
\end{table}

\begin{table}
\caption{Kinetic energies (in atomic units) of excited states of hydrogen like atoms. Z gives the 
atomic number of the atom and the excited state is such that the orbitals are occupied upto $n_1$, vacant 
from $n_1+1$ to $n_2$ and occupied again from $n_2+1$ to $n_3$ and the corresponding approximate kinetic 
energies. The latter are calculated by applying ground-state functionals $T^{(0)}$ and  $T^{(0)}+T^{(2)}$ of 
Equations~\ref{eq:tf} and \ref{eq:gea2} and the excited state functionals of Equations~\ref{eq:exke0} 
and~\ref{eq:exke2}. The corresponding errors are given below each number.}
\vspace{0.2in}
\begin{tabular}{p{1.0cm}p{1.0cm}p{1.0cm}p{1.0cm}p{1.4cm}p{2cm}p{2cm}p{2cm}p{2cm}}
\hline
Z& $n_{1}$   & $n_{2}$ & $n_{3}$ & $T^{(Exact)}$ &\hspace{0.2cm} $T^{(0)}$ &   $T^{(0)}+T^{(2)}$ & \hspace{0.2cm} $T^{*(0)}$ &  $T^{*(0)}$+$T^{*(2)}$\\
\hline
10&1&2&5&400&331.315&345.737&389.390&403.006\\
&&&&&(17.17)&(13.57)&(2.65)&(0.75)\\
15&2&4&6&900&700.795&737.727&873.249&907.667\\
&&&&&(22.13)&(18.03)&(2.97)&(0.85)\\
20&2&5&7&1600&1177.696&1249.214&1553.078&1620.010\\
&&&&&(26.39)&(21.92)&(2.93)&(1.25)\\
20&2&5&8&2000&1486.889&1558.398&1952.452&2019.384\\
&&&&&(25.66)&(22.08)&(2.38)&(0.97)\\
25&3&4&7&3750&3316.238&3437.872&3665.147&3779.984\\
&&&&&(11.57)&(8.32)&(2.26)&(0.80)\\
30&3&4&7&5400&4773.266&4960.075&5275.695&5456.973\\
&&&&&(11.61)&(8.15)&(2.30)&(1.06)\\
30&5&8&10&6300&5410.076&5597.432&6171.908&6353.568\\
&&&&&(14.13)&(11.15)&(2.03)&(0.85)\\
35&2&4&6&4900&3806.116&4071.146&4745.025&5017.298\\
&&&&&(22.32)&(16.91)&(3.16)&(2.39)\\
40&7&9&12&16000&14534.00&14904.93&15748.793&16139.184\\
&&&&&(9.16)&(6.84)&(1.57)&(0.87)\\
45&3&4&9&16200&14521.834&15010.222&15880.574&16426.318\\
&&&&&(10.36)&(7.34)&(1.97)&(1.40)\\

\hline
\end{tabular}
\end{table}

\begin{table}
\caption{Exact kinetic energies (in atomic units) of excited states of some atoms as obtained by solving the 
the Kohn-Sham equation with Gunnarsson-Lundquist parametrization of the LSD for exchange and correlation 
energy and the corresponding approximate kinetic energies. The latter are calculated by applying ground-state functionals 
$T^{(0)}$ and  $T^{(0)}+T^{(2)}$ of Equations~\ref{eq:tf} and \ref{eq:gea2} and the excited state functionals 
of Equations~\ref{eq:exke0} and~\ref{eq:exke2}. The corresponding errors are given below each number.}
\vspace{0.2in}
\begin{tabular}{p{4.20cm}p{2.0cm}p{2.0cm}p{2.0cm}p{2.0cm}p{2cm}}
\hline
Atom& $T^{(Exact)}$  & \hspace{0.2cm}  $T^{(0)}$ &   $T^{(0)}+T^{(2)}$ & \hspace{0.2cm}  $T^{*(0)}$ &  $T^{*(0)}$+$T^{*(2)}$\\
\hline

$Be(1s^{2}2s^{0}2p^{0}3s^{2})$ &13.768&12.278&13.768&12.459&13.945\\
&&(10.82)&(0.0)&(9.51)&(1.29)\\
$O(1s^{2}2s^{0}2p^{6})$ &73.094&64.154&70.068&67.545&73.704\\
&&(12.23)&(4.14)&(7.59)&(0.83)\\
$O(1s^{2}2s^{0}2p^{0}3s^{2}3p^{4})$ &65.764&56.967&63.516&59.834&66.291\\
&&(13.38)&(3.42)&(9.02)&(0.80)\\
$O(1s^{2}2s^{0}2p^{0}3s^{0}3p^{6})$ &65.506&56.344&62.815&59.885&66.313\\
&&(13.99)&(4.11)&(8.58)&(1.23)\\
$Ne(1s^{2}2s^{0}2p^{6}3s^{2})$ &124.508&109.521&118.891&116.152&125.947\\
&&(12.04)&(4.51)&(6.71)&(1.16)\\
$Ne(1s^{2}2s^{0}2p^{0}3s^{2}3p^{6})$ &109.241&93.430&103.889&99.675&109.920\\
&&(14.47)&(4.90)&(8.76)&(0.62)\\
$Mg(1s^{2}2s^{0}2p^{6}3s^{2}3p^{2})$ &191.942&169.083&182.740&180.095&194.392\\
&&(11.91)&(4.79)&(6.17)&(1.28)\\
$Ar(1s^{2}2s^{0}2p^{6}3s^{2}3p^{6}4s^{2})$ &501.507&443.200&474.770&474.671&507.648\\
&&(11.63)&(5.33)&(5.35)&(1.22)\\
\hline
\end{tabular}
\end{table}

\begin{table}
\caption{Exact kinetic energies (in atomic units) of pure excited states of some atoms (i.e. all the 
electrons have been excited) as obtained by solving the the Kohn-Sham equation with Gunnarsson-Lundquist 
parametrization of the LSD for exchange and correlation energy and the corresponding approximate kinetic 
energies. The latter are calculated by applying ground-state functionals $T^{(0)}$ and  $T^{(0)}+T^{(2)}$ of 
Equations~\ref{eq:tf} and \ref{eq:gea2} and the excited state functionals of Equations~\ref{eq:exke0} 
and~\ref{eq:exke2}. The corresponding errors are given below each number}
\vspace{0.2in}
\begin{tabular}{p{4.20cm}p{2.10cm}p{2.10cm}p{2.10cm}p{2.10cm}p{2.1cm}}
\hline
Atom& $T^{(Exact)}$  & \hspace{0.2cm} $T^{(0)}$ &   $T^{(0)}+T^{(2)}$ & \hspace{0.2cm} $T^{*(0)}$ &  $T^{*(0)}$+$T^{*(2)}$\\
\hline
$He(2s^{2})$ &0.736&0.181&0.263&0.575&0.595\\
&&(75.41)&(64.27)&(21.88)&(19.16)\\
$He(2s^{0}2p^{2})$ &0.676&0.292&0.315&0.614&0.606\\
&&(56.86)&(53.43)&(9.18)&(10.34)\\
$Be(2s^{2}2p^{2})$ &4.815&2.160&2.409&4.066&4.079\\
&&(55.14)&(49.97)&(15.57)&(15.30)\\
$Be(2p^{4})$ &4.565&2.219&2.370&4.337&4.277\\
&&(51.39)&(48.09)&(5.01)&(6.32)\\
$Be(3s^{2}3p^{2})$ &2.253&0.475&0.611&1.935&1.943\\
&&(78.94)&(72.89)&(14.11)&(13.78)\\
$O(2s^{2}2p^{6})$ &33.286&20.073&21.139&30.953&30.781\\
&&(39.70)&(36.49)&(7.01)&(7.53)\\
$O(3s^{2}3p^{6})$ &15.655&4.504&5.207&14.106&14.077\\
&&(71.23)&(66.74)&(9.89)&(10.08)\\
$Ne(2s^{2}2p^{6}3s^{2})$ &60.842&37.673&39.607&57.398&57.256\\
&&(38.08)&(34.90)&(5.66)&(5.89)\\
$Mg(2s^{2}2p^{6}3s^{2}3p^{2})$ &98.521&61.973&65.046&93.451&93.380\\
&&(37.10)&(33.98)&(5.15)&(5.22)\\
$Mg(2p^{6}3s^{2}3p^{4})$ &88.796&51.142&53.748&89.023&88.477\\
&&(42.40)&(39.47)&(0.25)&(0.36)\\
$Ar(2s^{2}2p^{6}3s^{2}3p^{6}4s^{2})$ &283.517&184.194&192.359&270.299&270.662\\
&&(35.03)&(32.15)&(4.66)&(4.53)\\
\hline
\end{tabular}
\end{table}

\begin{table}
\caption{Exact kinetic energies (in atomic units) of excited states of some atoms as obtained by solving the 
the Kohn-Sham equation with Gunnarsson-Lundquist parametrization of the LSD for exchange and correlation 
energy and the corresponding approximate kinetic energies obtained by applying the ground-state G\'{a}zquez functional 
 $T^{(0)g}$ of Equation \ref{eq:tgazq}   and the excited state functional $ T^{*g}$
of Equation \ref{eq:exgazq} . The corresponding errors are given below each number.}
\vspace{0.2in}
\begin{tabular}{p{4.20cm}p{2.0cm}p{2.0cm}p{2.0cm}p{2.0cm}p{2cm}}
\hline
Atom& $T^{(Exact)}$   &   $T^{(0)g}$  &  $ T^{*g}$\\
\hline
$Be(1s^{2}2s^{0}2p^{0}3s^{2})$ &13.768&14.472&14.282\\
&&(5.11)&(3.73)\\
$O(1s^{2}2s^{0}2p^{6})$ &73.094&69.760&73.151\\
&&(4.56)&(0.08)\\
$O(1s^{2}2s^{0}2p^{0}3s^{2}3p^{4})$ &65.764&73.210&70.675\\
&&(11.32)&(7.45)\\
$O(1s^{2}2s^{0}2p^{0}3s^{0}3p^{6})$ &65.506&72.756&69.242\\
&&(11.07)&(5.70)\\
$Ne(1s^{2}2s^{0}2p^{6}3s^{2})$ &124.508&118.552&124.722\\
&&(4.78)&(0.17)\\
$Ne(1s^{2}2s^{0}2p^{0}3s^{2}3p^{6})$ &109.241&123.323&118.816\\
&&(12.89)&(8.77)\\
$Mg(1s^{2}2s^{0}2p^{6}3s^{2}3p^{2})$ &191.942&182.374&192.450\\
&&(4.98)&(0.26)\\
$Ar(1s^{2}2s^{0}2p^{6}3s^{2}3p^{6}4s^{2})$ &501.507&480.681&508.486\\
&&(4.15)&(1.39)\\
\hline
\end{tabular}
\end{table}

\begin{table}
\caption{Exact kinetic energies (in atomic units) of pure excited states of some atoms (i.e. all the 
electrons have been excited) as obtained by solving the the Kohn-Sham equation with Gunnarsson-Lundquist 
parametrization of the LSD for exchange and correlation energy and the corresponding approximate kinetic 
energies obtained by applying the ground-state G\'{a}zquez functional 
 $T^{(0)g}$ of Equation \ref{eq:tgazq}   and the excited state functional $ T^{*g}$
of Equation \ref{eq:exgazq} . The corresponding errors are given below each number }
\vspace{0.2in}
\begin{tabular}{p{4.20cm}p{2.10cm}p{2.10cm}p{2.10cm}p{2.10cm}p{2.1cm}}
\hline
Atom& $T^{(Exact)}$   &   $T^{(0)g}$  &  $ T^{*g}$\\
\hline
$Be(2s^{2}2p^{2})$ &4.815&2.596&3.560\\
&&(46.08)&(26.07)\\
$Be(2p^{4})$ &4.565&1.721&3.303\\
&&(62.29)&(27.65)\\
$Be(3s^{2}3p^{2})$ &2.253&1.304&3.133\\
&&(42.12)&(39.07)\\
$O(2s^{2}2p^{6})$ &33.286&14.769&26.284\\
&&(55.63)&(21.04)\\
$O(3s^{2}3p^{6})$ &15.655&7.489&21.615\\
&&(52.16)&(38.07)\\
$Ne(2s^{2}2p^{6}3s^{2})$ &60.842&29.171&51.372\\
&&(52.05)&(15.57)\\
$Mg(2s^{2}2p^{6}3s^{2}3p^{2})$ &98.521&49.498&86.222\\
&&(49.76)&(12.48)\\
$Mg(2p^{6}3s^{2}3p^{4})$ &88.796&41.533&81.642\\
&&(53.23)&(8.06)\\
$Ar(2s^{2}2p^{6}3s^{2}3p^{6}4s^{2})$ &283.517&155.167&265.599\\
&&(45.27)&(6.32)\\
\hline
\end{tabular}
\end{table}

\end{document}